A Simulation of the Resilience of Cultural Diversity in the Face of Large-Scale Events

Ulloa, Roberto; Kacperski, Celina

Department of Computational Social Science, GESIS

University of Heidelberg

Abstract

Resilience and vulnerability of societies to large-scale events such as wars, climate catastrophes, immigration waves, and institutional collapses is not well-understood. Some societies prove more resilient, while others collapse. Diversity has been proposed in the literature as a potential cause of resilience in areas such as ecology, computer science and biology, and could also be a potential factor in sociocultural contexts. Using CulSim (Ulloa 2016), a cultural simulator that enables the study of cultural complexity and diversity based on Axelrod's (1997b) model of local convergence and global polarization and its extension including institutions (Ulloa et al 2016), the resilience of diverse societies and monocultural societies to nine different types of events is compared, while varying event sizes and distributions. Results show a strong positive relation between cultural diversity and resilience, in particular for events such as foreign invasions and settlements, and institutional damage and conversion.

**A Simulation of the Resilience of Cultural Diversity in the Face of Large-Scale Events**

Natural disasters, epidemics, wars, resulting immigration crises, and cultural clashes are currently common headlines of news reporting. Much has been written about the impact of these catastrophes on both individuals and societies (Alexander 2012; Drabek 2012; Drabek and McEntire 2003; Hoffman and Oliver-Smith 2002; Mach et al. 2019; Reyes 2006). From the perspective of archaeological theory, external factors were for a long time assumed to have little effect on development of cultural complexity over periods of time (for a discussion, see (Oliver-Smith 1996), and only over the past two decades, adaptation of societies to external events and their resilience has received a larger share of attention (Adger et al. 2005; Berkes and Ross 2013; Cutter et al. 2008; Gerstenberger and Nusser 2015; Delle Fave 2014). One focus has been, of course, the climate crisis (Adger et al. 2009; Berrang-Ford et al. 2019), another the flexibility of societies and its social context (Blaikie et al. 2014; Cutter et al. 2008; Oliver-Smith 1996). The resilience of societies - how they respond and recover - has been often studied with the aim to better provide disaster relief and improve response mechanisms in cases of large-scale events (Cutter, Boruff, and Shirley 2003; Eriksson et al. 2015; Janssen et al. 2006; Montz, Tobin, and Hagelman 2017). A pertinent question is then how such resilience emerges and is upkept, and how it can be strengthened. Previous studies in social resilience have found increases through improvements in communications, risk awareness, and preparedness (Paton and Johnston 2017; Paton, Smith, and Violanti 2000). In ecology, starting from a much more basic assumption, ecological resilience has been shown to benefit simply from ecological diversity (Adger 2000; 2006; Adger et al. 2005; Holling 1973), and the diversity of human activity around the ecosystem (Leslie and McCabe 2013). Further, arguments have been made that non-diverse livelihoods are more vulnerable to economical and natural disasters (Marschke and Berkes 2006; Martin and Lorenzen 2016; Thulstrup 2015).

The association of diversity as a mechanism of defense is supported by evidence in other research fields, such as economics (Brown and Greenbaum 2017), computer security (Borbor et al. 2019), and biology: ecosystems, gene pools, immune systems (Xu, Böttcher, and Chou 2020). Moreover, similarities between biology and culture have already been highlighted in the literature, e.g., cultural evolution (Boyd and Richerson 1985). As diversity emerges and reaches equilibria in both cultural and biological system, following this association, we propose that diversity can play a defense mechanism against perturbations in cultural systems as well.

**Events and cultural resilience**

In this paper, events are defined as any occurrences that have a sudden impact on the simulated society, either to its members or its institutions, similar to the definition of (Torry et al. 1979) as "events that lead to public and private facilities being unable to provide essential social and economic services without extensive replacement" (1979, 517) . Catastrophic events differ by type and amount of damage that they inflict, e.g. rates of mortality, destruction or abandonment. War and invasions, environmental disasters (earthquakes, droughts, hurricanes), internal structural failures (such as a genocides or economic breakdowns) and human-shaped disasters (such as land degradations, or nuclear accidents) are typical examples. Other large-scale events include settlements and major immigration waves.

Grattan and Torrence (2003), speaking of natural disasters, proposed that they are definite triggers of cultural change, both in the past and in modern times, and the magnitude of this change can be a function of, among other things, resilience. In the context of the studies presented here, cultural groups and their resilience will be the one of the main variables of interest. Culture is here defined as "information which is transmitted between individuals in a social manner" (Ulloa, Kacperski, and Sancho 2016, 2); the process of this cultural transmission is in the literature known as social influence (Festinger, Back, and Schachter 1950). Cultural resilience is used in the vein of ecological resilience, as a "measure of the persistence of systems and their ability to absorb change and disturbance and still maintain the same relationships between populations or state variables" (Holling 1973, 14). Thus, we will measure the ability of a society to retain its cultural composition, both cultural content and geographical distribution across the affected population. The focus is on long-term effects of events, ascertaining the impact of the event when a new societal equilibrium has been reached.

Historical and sociological analyses of example societies yielded no clear answer how large-scale events trigger cultural change across societies, or what makes some social systems more resilient than others (Masson, Hare, and Lope 2010; Morris 2010; Nichols and Weber 2010; Oliver-Smith 1986; Sheets and Grayson 1979). For example, in the 17[th] century, European society allegedly recovered almost unscathed from volcanic eruptions leading to widespread catastrophic toxicity; one might have expected it to long-term affect agricultural production and thus European cultural patterns (Brayshay and Grattan 1999). On the other hand, geographical movement and cultural adaptation have been observed in the Moche society in Peru following a longer-term exposure to a series of environmental catastrophes such as massive flooding, erosion and mass wasting (Kornbacher 2003). In Syria, wars, raids and environmental stresses resulted in multiple instances of collapse and cultural regeneration in the Euphrates Valley and Jabbul Plain in the Bronze Age. Their diverse cultural composition remained largely intact, arguably due the capacity of settlements to make autonomous decisions and their economic advantages (Cooper 2010; Nichols and Weber 2010). Computer simulations have shown that geography plays a role (Greig 2002; Parisi, Cecconi, and Natale 2003): elevated interior areas of the Yucatán Peninsula, when simulated, were more susceptible to system collapse and less suitable for resilient recovery than adjacent lower-lying areas (Dunning et al. 2002). Related to this, the distance between the groups' centers (e.g. cities or tribes) could serve to increase resilience, as would be the case for the Tarahumaras (Sheridan and Naylor 1979), who have proven resilient and been able to sustain their culture despite their more homogenous society. Finally, it has been proposed that institutions impact cultural development, in particular resilience and vulnerability of cultures, more than any other factor (Schwartz and Nichols 2010). Interestingly, political and cultural disintegration do not always happen in parallel – culturally stable societies have been known to rebuild almost identical political systems after a collapse, while others establish novel systems; and culturally transformed societies can keep old political systems intact (Schwartz and Nichols 2010). Thus, it can be hypothesized that the way in which institutions are targeted by different events play a role – by their connection to their members, or by their content itself.

To summarize, three major areas of investigation emerge from the analysis of change events in the context of cultural diversity and resilience. For one, the investigation of the magnitude of event is of importance. Secondly, the investigation of different types of events is necessary, with the literature considering offering diverse classifications and frameworks to assess their impact (Quarantelli, Boin, and Lagadec 2018; Diamond 2011; Webster 2002, 327–29). Based on these, a first definition of a list of events has been proposed (Ulloa 2016). Thirdly, monocultural and multicultural diverse societies will be compared in terms of their resilience in the face of the different type and intensities of events. We hold the hypothesis that multicultural societies, i.e. those consisting of multiple cultural groups, will show higher resilience than monocultural societies. Moreover, for the multicultural societies, we will perform explorative analyses regarding the effect of preexisting diversity on resilience, as well as its interactions with the event parameters.

In the next section, we will review the ideas behind the institutional model (Ulloa, Kacperski, and Sancho 2016) that we used for the analysis for these three lines of investigation.

**Models of social influence and institutional model**

A major difficulty in studying the impact of large-scale perturbations is methodological - they cannot be experimentally manipulated, causation is difficult to establish, and the relationship of cultural resilience (or, its counterpart, cultural vulnerability) and change events is often seen as co-occurring but not necessarily interacting (Grattan and Torrence 2003). Computer simulations can circumvent these issues, and so agent-based models have become a popular tool for empirical testing of realistic concepts, and have proven to be a successful paradigm for modeling complex systems (Niazi and Hussain 2013). Initial agent-based models were common in economy and biology (Föllmer 1974; Wolfram 1983); in the social sciences, they have been used to study social systems (Epstein 2012; Epstein and Axtell 1996), to explain spatiotemporal history (Dean et al. 2000), and to predict cooperation and competition in societies (Axelrod and Hamilton 1981; Axelrod 1997a). Dissemination of culture through social influence (Axelrod 1997b), including the impact of institutions on cultural diversity (Ulloa, Kacperski, and Sancho 2016) has been examined with cultural simulations that are termed *artificial societies* (Epstein and Axtell 1996). In Axelrod's seminal research, cultural features (e.g. music, food) are introduced as a list of categories (e.g. rock music, coffee), with the traits on those categories represented by numerical values. The similarity of agents on multiple of these features is taken as a representation of social similarity. Agents then undergo instances of social influence, following the concept of homophily, the principle that "like attracts like" (McPherson, Smith-Lovin, and Cook 2001). Outcome states of the model can be diversity, i.e. a society with many different clusters of cultures, one big uniform culture (monoculture), or all individuals possessing a culture that is different from all their neighbors, i.e. anomie (Durkheim 1951). These models can gather support for causal hypotheses regarding stability of social systems following disturbances: on one hand, to explore which features increase cultural resilience, on the other hand, to investigate which are the most adverse events to cultural diversity.

Axelrod's model has been extended manifold: most notably, it has been shown that the original model is very susceptible to the introduction of perturbations such as mutations (i.e. a sudden innovations or change of mind in an agent) or selection errors (i.e. a misperception of the similarity of other agents) (Flache and Macy 2011; Klemm et al. 2003). Many theoretical contributions have been made to stabilize the model, such as the inclusion of institutions that emerge from human interactions (Ulloa, Kacperski, and Sancho 2016). The role institutions play in the social domain can be understood to mostly depend on explicit and implicit rules of behavior, including ideologies, sociopolitical institutions and beliefs as well as physical spaces such as libraries or school systems (Hodgson 2006; Knight 1992). In the institutional model that we use for the here conducted research, institutions have been conceptualized in terms of repositories (cultural information centers). These institutions influence individuals and their behavior, while being influenced by individuals in return (Ulloa, Kacperski, and Sancho 2016). Experimental data on the impact of institutions on culture and its underlying processes of social influence are rare. To the best of our knowledge, just a few agent-based models exist, studying authoritarian regimes (institutions), integration of information repositories, institutional effectiveness (Bhavnani 2003; Makowsky and Rubin 2013; Suarez and Sancho 2010) and mass media influence on cultural diversity (Shibanai, Yasuno, and Ishiguro 2001). Ulloa, Kacperski and Sancho (2016), in a first direct examination via agent based model, found that institutional influence (i.e. the pressure an institution puts on its members to prevent social influence) increased and preserved cultural diversity - and that there is an important interaction of a top-down and bottom-up information flow. While top-down processes of institutional information dissemination, such as propaganda, increased diversity, bottom-up processes typically common in democracies, such as voting or referenda, promoted convergence towards homogeneity (i.e. more globalization).

CulSim (Ulloa, 2016) is an agent-based model software that provides several of the above-mentioned cultural models (Axelrod, 1997b; Flache & Macy, 2011; Ulloa et al., 2016) and the possibility to simulate events such as wars, mass migration and institutional collapse. Of the models included in the software, we selected the institutional model (Ulloa, Kacperski, and Sancho 2016), which supports all possible events implemented in CulSim, including those that affect institutions.

## Methodology

### Events model

In the simulation, each individual of a society (agent) is a cell of a grid. The two images of Figure 1 contain 25 individuals (a 5x5 grid) and 4 cultures represented by different colors. A cultural group is defined as those individuals that are adjacent to each other and that also have the same cultural traits on each of the possible cultural features - the two green agents (cells with arrows pointing out their cultural vectors) belong to the same culture (green), because their two features (music and sport) both hold the same trait each (jazz and tennis respectively).

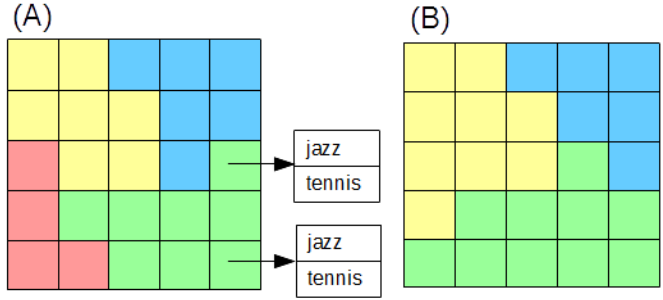

*Figure 1. Hypothetical example of two cultural spaces of states of a simulation at different times.* Each of the 25 cells (5x5 grid) represent an individual agent, with colored areas representing cultural groups. As an illustration, the two possible cultural features are music (jazz) and sport (tennis). The left image (A) presents the state of the simulation at time t, and the right image at time t + 100.

The cultural groups in Figure 1. Hypothetical example of two cultural spaces of states of a simulation at different times can be characterized by three descriptors: the number of individuals that belong to it (*size*), the *place* they occupy on the grid (*position*), and the cultural traits that they contain (*content*). These criteria are used to calculate the similarity between two *states* of the simulation. A *state* of the simulation is a snapshot of the cultural composition at any given time.

For example, the left side image (A) of Figure 1. Hypothetical example of two cultural spaces of states of a simulation at different times represents the cultural composition of the simulation at time *t*, whereas the right-side image (B) represents the state at time *t+100*. In the simulation, time is controlled via iterations. During an iteration, the agents may interact with their neighbors (i.e. the adjacent cells) and share traits according to a probability that is given by their homophily, calculated as follows:

(1)     $H(x, y) = \frac{1}{F} \sum_{f=1}^{F} \delta(x_f, y_f)$

Here, *F* is the number of features, and *δ(i,j)* refers to the Krockener delta; i.e. $\delta(i, j) = 1 \; if \; i = j, 0 \; otherwise$. These interactions will change the cultural composition of the system.

Starting from a random assignment of traits among the agents, homogeneous groups emerge from the system.

Given two states at times $t_1$ and $t_2$, e.g. A and B of Figure 1, each cultural group in A can be compared to the most similar one in B in terms of the three descriptors (*size, position* and *content*). These descriptors can be used to find the most similar cultural group. The similarity between two cultural groups *a*, which belongs to the state A, and *b*, which belongs to the state B, is calculated for each individual descriptor as follows:

- $Sim_{size}(a, b) = 1 - |a_{size} - b_{size}|$, where $g_{size}$ is the size of group g

- $Sim_{position}(a, b) = 1 - \sqrt{\left(\frac{a_x - b_x}{N}\right)^2 + \left(\frac{a_y - b_y}{N}\right)^2}$, where $g_x$ and $g_y$ represents the central coordinate x and y of group g, and N the totals of columns (or rows) of the grid.

- $Sim_{content}(a, b) = \frac{1}{F}\sum_{f=1}^{F} \delta\left(a_f, b_f\right)$, where $\delta(i,j) = \begin{cases} 1, & i = j \\ 0, & i \neq j \end{cases}$, $g_f$ represent the trait for the cultural $f_{th}$ feature of cultural group *g*, and *F* represent the total number of features

Then, the aggregated similarity between *a* and *b* is calculated by multiplying the three previous similarities:

(2)     $Sim_{group}(a, b) = Sim_{size}(a, b) * Sim_{content}(a, b) * Sim_{position}(a, b)$

It is then possible to establish a similarity measure by finding the most similar group in B for all of the groups in A, and vice versa, and then averaging the similarity of the pairs:

$$Sim_{st}(A, B) = \frac{\sum_{a \in A}^{A} min(\{Sim_{group}(a,b):b\ in\ B\}) + \sum_{b \in B}^{B} min(\{Sim_{group}(b,a):a\ in\ A\})}{\|A\| + \|B\|}$$

Notice that in the numerator of the formula, the comparison is done from A to B and from B to A. This is necessary because it is possible to have an unmatched number of cultural groups from one state to another, for example, in Figure 1, the bottom left group (red) of state A extinguished and does not exist in the state B. By considering a comparison in both directions, all cultural groups get an opportunity to compare themselves to one another. Also, a cultural group qualifies as such only, if it contains a minimum of three members, as suggested by Flache & Macy (2011), and based on the idea that triad social interactions are fundamental for social consensus (Simmel 1950).

Figure 2 shows two examples of the graphical interface of CulSim – on the left, it shows a scenario starting with 14 cultures and ending with 99, at almost 60% similarity; on the right, it shows a scenario starting with 24 cultures and ending with 153 at a 75% similarity.

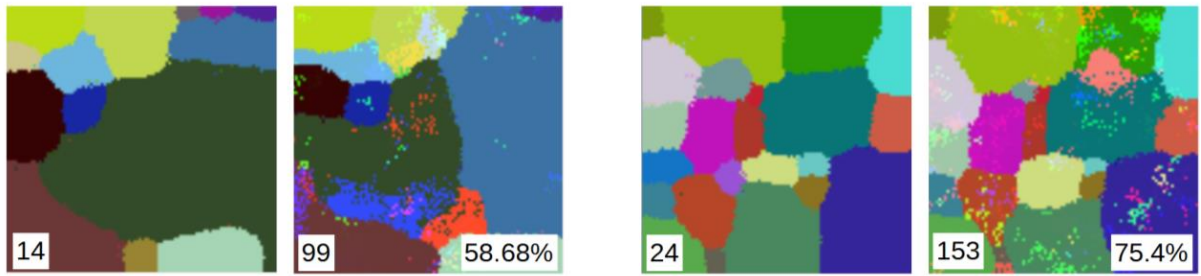

*Figure 2. Example of the CulSum interface displaying two instances of a possible society with different cultural groups and the effect of an event.*

The model that we use in this paper also includes institutions, which influence the cultural traits that an individual may adopt when an interaction between agents occurs. Each agent can affiliate to an institution, and an institution is generally composed by a collection of several agents (see Figure 3 for an idealized example, in which all agents of their respective cultural groups are affiliated to the same institutions).

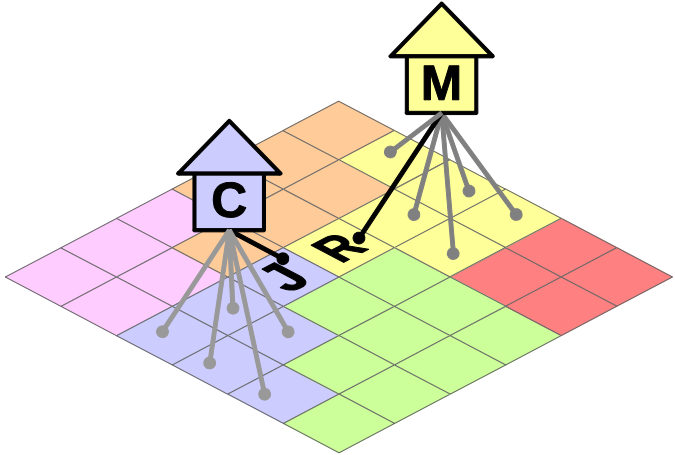

*Figure 3. Institutional model. Reproduced with permission from Ulloa et al. (2016). "J" and "R" represent agents, "M" and "C" their respective institutions, Agents form their cultural regions (e.g. yellow, blue) and are connected to their respective institutions (also colored yellow and blue).*

The influence that the institutions exert over their affiliates is given by two components: the *institutional influence*, how much can an institution prevent a change towards a neighbor's trait, and *agent's loyalty,* how likely is it that an agent would remain in their current institution after accepting a neighbor's trait. Equation (3) expresses how the institutional influence is calculated and controlled by the $a$ parameter; $i_x$ is the institution to which agent $x$ is affiliated, and $H$ was defined in equation (1), $H$ can also be used to calculate the similarity between an agent and institution.

(3)     $Inf(x,y) = \frac{\alpha \cdot H(x, i_x)}{(1-\alpha) \cdot H(x,y) + \alpha \cdot H(x, i_x)}$ , where $\alpha \; \epsilon \; [0,1]$

Equation (4) expresses how the agent's loyalty is calculated and controlled by the $\alpha'$ parameter (see also Ulloa et al., 2016).

(4)     $Loy(x,y) = \frac{\alpha' \cdot H(x, i_x)}{(1-\alpha') \cdot H(x, i_y) + \alpha' \cdot H(x, i_y)}$ , where $\alpha' \; \epsilon \; [0,1]$

For reference, Table 1 summarizes the parameters of the model described so far.

*Table 1. Parameters of the institutional model.*

| Parameter | Description |
|---|---|
| Size (N) | Number of agents given by (R)ows x (C)olumns in the grid |
| Features (F) | Number of topics of opinion |
| Traits (T) | Possible values of each topic |
| Mutation (M) | Probability of randomly changing a trait |
| Selection Error (E) | Probability of making a homophilic error, reverting the result of $H(x, y)$ |
| Vicinity (V) | Manhattan distance that defines an agent's neighbors. |
| α and α´ | Control the institutional influence (α) and agent loyalty (α´) |

**Overview over events**

Apart from configuration parameters of the institutional model (Table 1), CulSim provides nine events that were designed exhaustively to affect the information in the system (Ulloa 2016). They can potentially be combined to approximate historical events, however, in this paper, we analyze each of these events in isolation as a baseline for future studies that intend to analyze multifactorial historical events. Table 2 lists examples for isolated events from CulSim; these are purely illustrative for their most salient characteristics - most real live examples involve a combination of events.

The settlement and immigration events imply the introduction of new agents which is not possible in a fixed grid, given that the maximum capacity of the model is limited to the grid size (i.e. the number of agents is given by the number of rows times the number of columns); therefore, the new agents replace existing ones. As we are primarily interested in the change of cultural composition in the grid, this is still captured adequately, and future models can adapt the behavior of individual agents to better suit realistic interpretations.

*Table 2. Overview over events. Column 1 names the event, column 2 indicates the main target of the event: individuals, institutions or institutional traits. Columns 3 and 4 provide description and examples respectively. Note that by attacking individuals, damage to institutions can potentially occur (e.g. apostasy).*

| Event | Main Target | Description | Event example |
|---|---|---|---|
| **Decimation** | Individuals | A percentage of the population is killed. Dead individuals are represented as agents with empty traits. | Pandemics, natural disasters such as tornados |
| **Settlement** | Individuals | A percentage of existing individuals are replaced by settlers, represented as agents with foreign traits, all linked to one institution with foreign traits. | Restructuring of post-war regions through forced resettlement |
| **Outsiders** | Individuals | A percentage of existing individuals are replaced by outsiders, represented as agents with foreign traits who are not linked to any institution. | Institution-less immigrants, raids |
| **Apostasy** | Individuals | A percentage of the population become apostates. Apostates are agents not linked to an institution. | System disillusionment or pre-revolutionary states |
| **Institutional destruction** | Institutions | A percentage of institutions are destroyed. Their agents become not linked to any institution. | Post-revolutionary states, statelessness; susceptibility to new institutions |
| **Partial content removal** | Institutional traits | A percentage of institutional traits are removed from the existent institutions. | Destruction of educational and value systems, for example of libraries and museums |
| **Full content removal** | Institutional traits | All institutional traits from a percentage of institutions are removed. | |
| **Partial conversion** | Institutional traits | A percentage of institutional traits are converted to foreign traits. | Replacement of educational and value systems with foreign equivalents, e.g. forced religious conversions |
| **Full conversion** | Institutional traits | All institutional traits from a percentage of institutions are converted to foreign traits. | |

All the events in Table 2 can be configured. First, a distribution of the event is assigned, choosing from two available possibilities: uniform and centered. The uniform distribution equally affects all entities of the simulation. The centered distribution uses a normal distribution to concentrate the event in a given location.

Secondly, the size of the event is assigned, by choosing an expected percentage of entities (individuals, institutions or traits depending on the event) that will be affected by the event. CulSim will select a probability that will maximize the likelihood of the chosen percentage. For example, if the percentage of a uniformly distributed decimation is 20%, each agent is given a 0.2 probability of being removed from the grid. A similar logic is applied for normally distributed decimations; however, probabilities are selected based on their proximity to the center of the event. For both types of distributions, an expected percentage, referred to as the event size, is the parameter.

Third, for normally distributed probabilities only, two other parameters must be assigned: the center of the event, i.e. a coordinate that will receive the highest probability of the normal distribution, and the ceiling of the distribution, i.e. the value of the highest probability assigned to the cell in the center of event.

While CulSim provides multiple response variables that can be used to analyze the effects of events, the response variable used in the present study is the similarity between the state before the event (s_bef), and the state 100,000 iterations after the event (s_aft), i.e. Sim_states ($s\_bef,s\_aft$). For the experiments, this variable will be called similarity and it will be used as a measurement of cultural resilience.

**Selected scenarios**

Five factors will be manipulated to test differences of cultural composition pre and post-event. These factors are summarized in Table 3 below. For interpretation of results, event-related factors will be the most interesting, as well as number of cultures pre-event.

*Table 3. Overview over simulation factors. Column 1 names the factor, column 2 describes the type of value of the factor, column 3 explains the factor with examples.*

| Factors | Applied as | Description and Example |
|---|---|---|
| Type of event | Categorical | For a list of all events, see Table 4 |
| Size of event | Percentage | Extent to which an event targets society agents or institutions, e.g., decimation of 100% means all agents are removed. |
| Distribution of event | Uniform vs. normal | Spread of events' effects across society, e.g., uniform distribution affects all areas equally likely, a normal distribution has an "epicenter". |
| Population size | (R)ows x (C)olums | Number of agents, see Table 4. |
| Number of cultures | Numerical | Number of cultures that exist in a stable formation after 100,000 iteration of the simulation pre-event. For example, 1 as a monoculture, 6 in a small-scale society and 44 in a large-scale society. |
| Institutional influence | Percentage | The parameter $\alpha$ that controls the institutional influence, see Table 4 |

A subset of scenarios has been chosen that were deemed appropriate for first testing, based on previous results (Ulloa, Kacperski, and Sancho 2016), see Table 4.

*Table 4. Selected scenarios are characterized by an identifier (1st column), main parameters (columns 2-4), and the average number of emerged groups of the 24 simulation runs after 1,000,000 iterations*

| Scenario (S) | Population (NXN) | Radius (R) | Institutional influence (I) | Emerged Groups (G) |
|---|---|---|---|---|
| A | 32x32 | 6 | 0.85 | 5.9 |
| B | 32x32 | 3 | 0.85 | 18.2 |
| C | 100x100 | 6 | 0.80 | 30.8 |
| D | 100x100 | 6 | 0.85 | 44.5 |

In terms of notation, the scenarios of Table 4 are going to be identified as follows: S(G): NxN/R/I. The meaning of initials S, N, I and R, are given in parentheses in the header column of Table 5. The letter G represents the average number of cultural groups generated by the scenario.

Values of the parameters listed in Table 4 were chosen based on prior literature. Scenarios A and C were previously explored (Ulloa, Kacperski, and Sancho 2016) . B and D are variants of A and C respectively, and were chosen to obtain variety for the factor "number of cultural groups, pre-event" - for example, larger neighborhood interactions decrease the number of cultural groups in simple versions of the here presented model (Greig 2002) , and so do smaller levels of institutional influence (Ulloa, Kacperski, and Sancho 2016). These scenarios will provide a basis for making generalized statements about the impact of the events listed in Table 4.

All other parameters used in the current simulation are held constant, consistent with previous literature (Axelrod 1997b; Flache and Macy 2011; Ulloa, Kacperski, and Sancho 2016), agents hold 5 features (F) and 15 possible traits (T); agent's loyalty ($\alpha'$) is set at 0.5; both noise sources, mutation and selection error, are set at 0.001; the number of iterations before the event are set at 1,000,000 (i.e. the event always occurs at iteration 1,000,000) and the similarity is calculated at 100,000 iterations after the event.

Since the simulation is non-deterministic, 24 repetitions with each scenario are run. To avoid variance disturbances due to different initial conditions, the 4 scenarios (without events) and 24 repetitions are run until 1,000,000 iterations are reached (i.e. before the event). At this point, the states of the 96 repetitions are stored and loaded to execute each of the events. Therefore, all events will be executed on the same 24 conditions per scenario. A variety of cultural groups will exist pre-event, all with different institutional arrangements (Ulloa,

Kacperski, and Sancho 2016); the last column of Table 4 shows the average number of groups that emerged after 1,000,000 iterations on the 24 repetitions. We consider all these "diverse scenarios", as they contain several groups. Correspondingly, we generate a monoculture society scenario for each diverse scenario of Table 4. While maintaining the parameters of the simulation constant, we create a state in which all individuals share the same cultural traits and are assigned to the same institution. Note that the notation S(G): NxN/R/I will also be used for the monoculture scenarios with the particularity that the number of groups (G) is always 1; for example, the monoculture scenario A(**1**)=32x32/6/.85 corresponds to the diverse scenario A(**5.9**)=32x32/6/.85.   As event effects are applied, resilience of societies with cultural diversity will be compared to resilience of monocultural societies.

## Results

In the following subsections, we present the results obtained from the scenarios described in the methodology. Each subsection contains graphs in two columns: the left showcases results from uniformly distributed events and the right from normally distributed events. Each graph contains 8 series: 4 correspond to diverse scenarios (red), and 4 to monoculture scenarios (blue). The dependent variable is always the long-term resilience of the society's cultural composition, i.e. similarity of the resulting state 100,000 iterations after the event occurred, to the state just before the event. As a reference, when the similarity is above 0.8 the states are visually very similar (the similarity is only high if the size, content and location are similar, see Equation (2)), whereas similarities below 0.2 represent societies that have changed almost entirely.

### Decimation

Figure 4 presents the effects of uniform and normally distributed *decimation*.

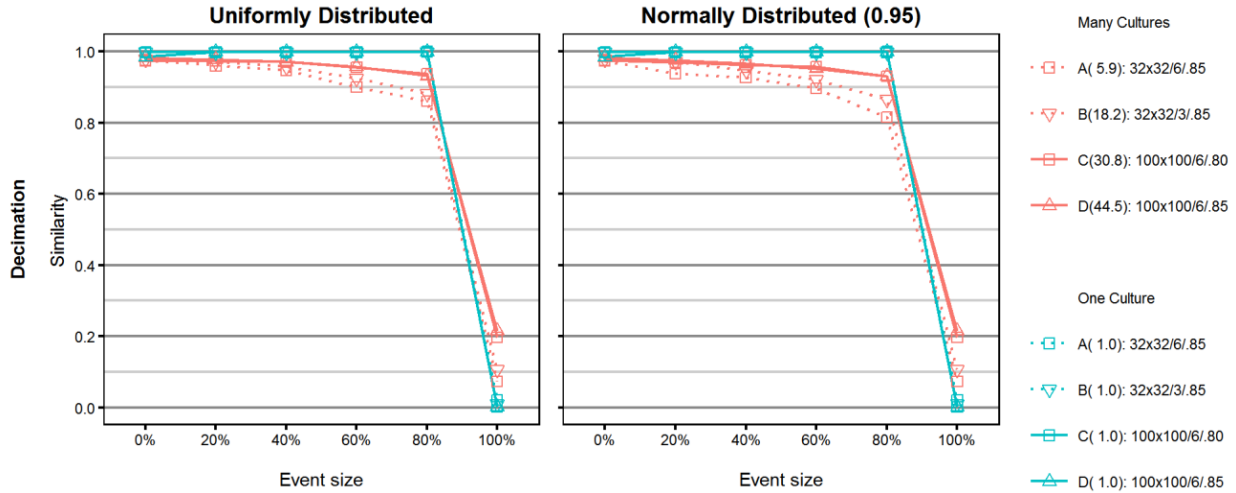

*Figure 4. Effects of decimation.* Legend: S(G): NxN/R/I where S is the identifier; G is the average number of cultural groups; NxN is the number of rows and column; R is the distance for neighborhood interaction; I is the level of institutional influence. Y-axis: similarity between the state just before the event (1,000,000th iteration), and 100,000 iterations after the event. X-axis: event size as a percentage of the affected agents.

Following the X-axis with increasing event size, we can observe that the similarity of pre- and post-event state stays high for all scenarios until the event size is at least 80%. For reference, the European black death of the 14th century has been estimated to have eradicated more than 50% of the population in urban areas (Christakos 2005). Whether diverse or homogenous, the cultural compositions of the simulated societies are highly resilient to decimation. Comparing monoculture (blue) with diverse societies (red), the latter do not recover as completely at smaller event sizes, independently of the distribution of the event: some changes introduced by the event persist.

**Foreigners**

In Figure 5, we distinguish between two different types of foreigner events: *settlement* (top row, agents that all have a shared institution in common) and *outsiders* (bottom row, agents don't belong to any institution) – in both cases, the event size is represented as a percentage of the territory being occupied by a group of foreigners.

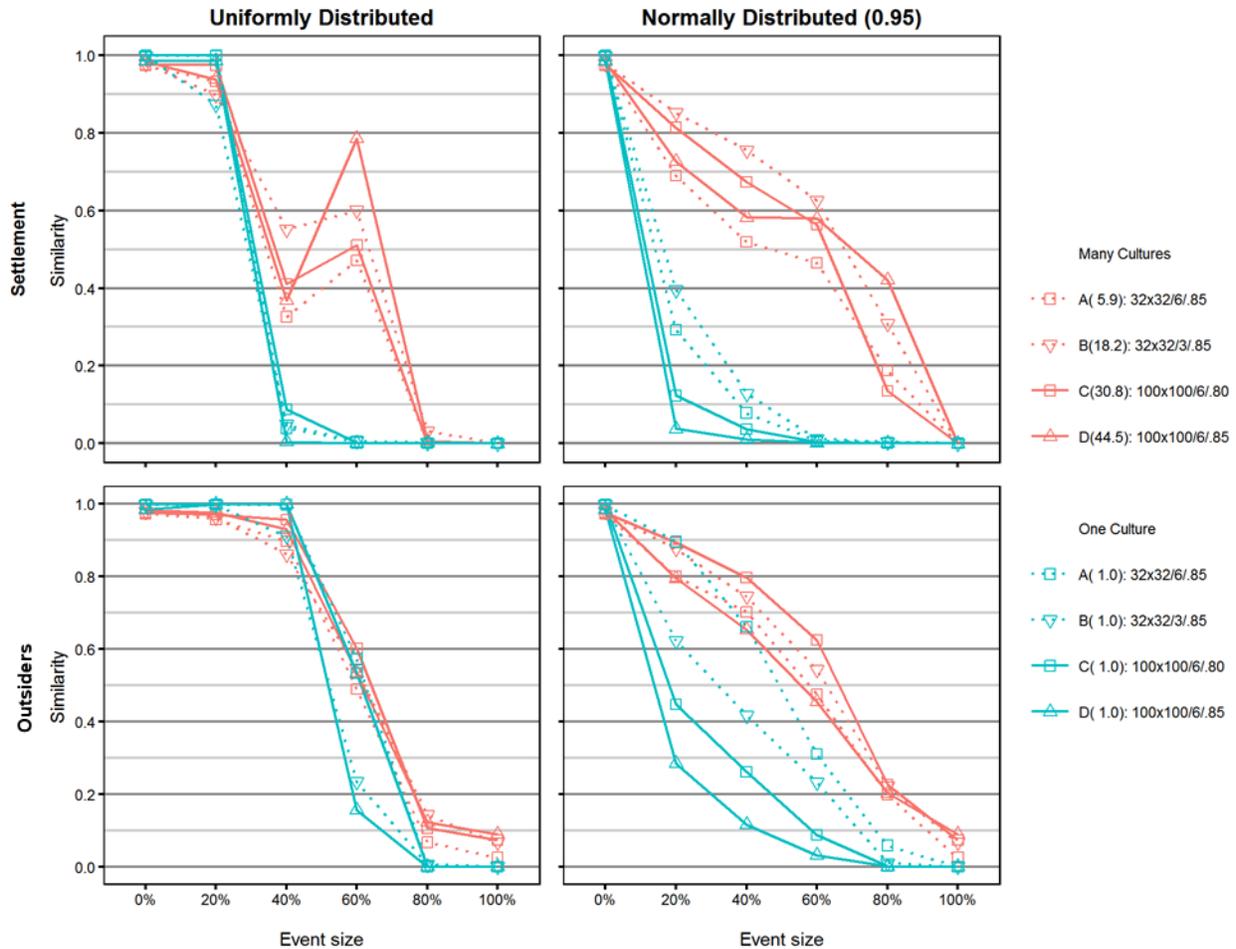

*Figure 5. Effects of settlements and outsiders.* Legend: S(G): NxN/R/I where S is the identifier; G is the average number of cultural groups; NxN is the number of rows and column; R is the distance for neighborhood interaction; I is the level of institutional influence. Y-axis: similarity between the state just before the event (1,000,000[th] iteration), and 100,000 iterations after the event. X-axis: event size as a percentage of the affected agents.

Both events have a high impact on cultural similarity, much higher than *decimation* alone did: the introduction of a *settlement* of even a small size decreases the similarity for all scenarios. For reference, migration numbers between 1% and 10% have been documented across the EU in the past years (Eurostat 2020; UNDESA 2015), though countries such as the United Arab

Emirates or Kuwait report numbers of immigrants at over 70% percent ("World Migration Report 2020" 2019)

    *Settlement* has a stronger impact than *outsiders* (the downward gradient is higher), meaning it takes larger numbers of institution-less agents to change the cultural composition of a society in the long-term. The settler agents' institutional allegiance stabilizes their cultural group, so that the resulting cultural composition reaches an equilibrium that persist after the long after the event. In contrast, outsider agents oftentimes adopt existing institutions. However, at high numbers, outsider agents are more likely to be "neighbors", and then are more likely to build their own institution (i.e. they end up acting in a similar fashion as in the *settlement* scenario).

    The distribution of the events plays a major role. The monoculture societies' culture only survives under smaller sizes of uniformly distributed events, whereas they are critically affected by the lowest tested size (20%) of a normally distributed event. The simulated diverse societies display a stronger resilience to normalized distributed event (red lines are often above the blue lines in the right graphs), whereas the resilience for uniform events is similar to the one of the monoculture societies (albeit higher for larger event sizes in settlement).

    We don't find a clear effect for the number of preexisting cultures affecting resilience; this would reflect in the red triangular shapes showing a higher similarity than the square shapes consistently; a trend can be spotted following that higher preexisting levels of diversity make a society slightly more resilient when settlement occurs at very large sized effects (i.e. over 60% in the uniform scenario and at exactly 80% in the normally distributed scenario).

**Institutional Damage**

    Figure 6 shows the effects of two different types of institutional structural damage, *apostasy* (top; abandonment of institutions by agents) and *institutional destruction* (bottom;

destruction of institutions). In these scenarios, agents themselves are not directly targeted by external events; either the connection between agents and their institution is removed, or the institution itself is removed.

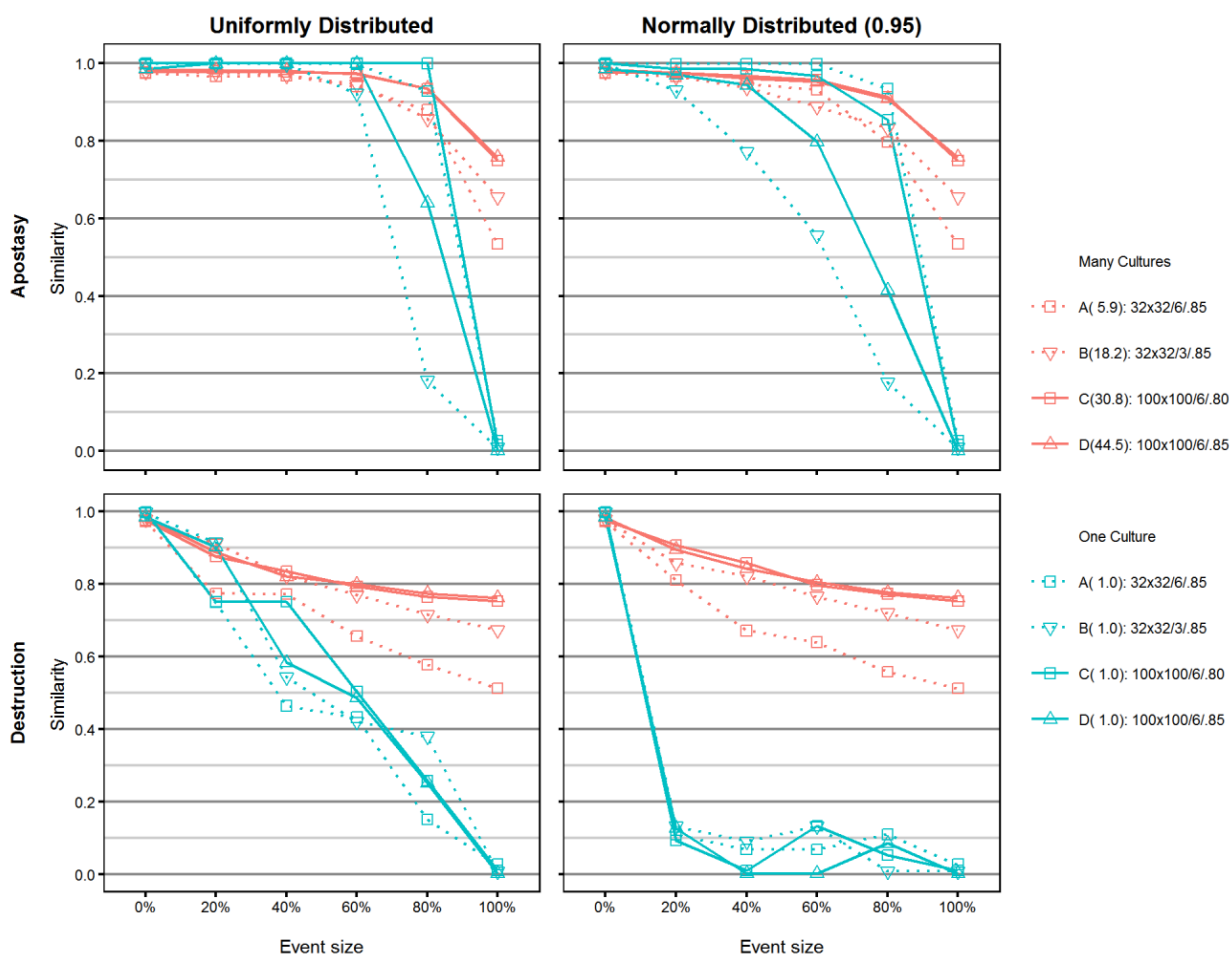

*Figure 6. Effects of apostasy and institutional destruction.* Legend: S(G): NxN/R/I where S is the identifier; G is the average number of cultural groups; NxN is the number of rows and column; R is the distance for neighborhood interaction; I is the level of institutional influence. Y-axis: similarity between the state just before the event (1,000,000[th] iteration), and 100,000 iterations after the event. X-axis: event size as a percentage of the affected agents.

Apostasy changes the composition of the simulated societies only at very high levels, similar to results we find in the decimation scenario. In contrast, institutional destruction interacts with the distribution of the event and with the diversity of a society pre-event. In this case, diverse societies prove to be quite resilient to even complete institutional structural damage; i.e. at the rate of 100% damage, many aspects of the pre-event cultural composition are preserved. Moreover, among the diverse scenarios in a 32x32 grid, there is a positive correlation between diversity and resilience, i.e. the more cultural groups exist in the scenario (pre-event), the higher the similarity (red triangles consistently stay above the red squares in the figure). For a 100x100 scenario, the preexisting number of cultural groups seems not to play a major role anymore; this might be due to the fact that they start with a higher number of cultural groups to begin with.

The culturally homogenous simulated societies tolerate high rates of uniform apostasy but swiftly collapse at high rates; for normally distributed apostasy, their cultural similarity post-event varies greatly depending on neighborhood interaction distance and very small differences in institutional influence. Finally, the simulated homogenous societies are particularly vulnerable to normally distributed institutional destruction - at event sizes of even just 20%, the cultural similarity pre- and post-event drops to around 15%.

**Institutional Content Removal**

Figure 7 shows the effects of uniform and normally distributed *institutional content removal* on cultural composition of the simulated societies. *Partial removal* (top row) and *full content removal* (bottom row) are explored.

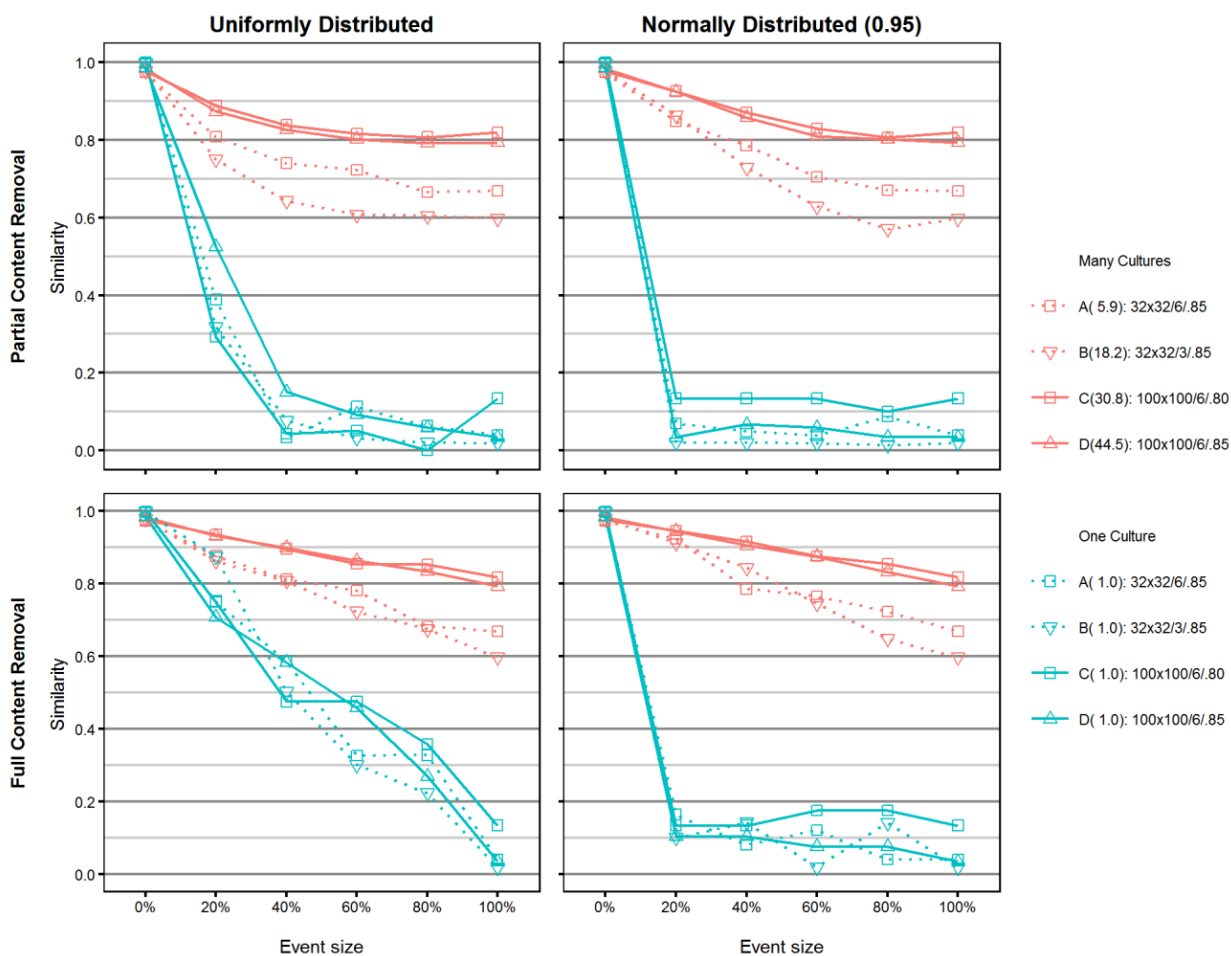

*Figure 7. Effects of partial and full institutional content removal.* Legend: S(G): NxN/R/I where S is the identifier; G is the average number of cultural groups; NxN is the number of rows and column; R is the distance for neighborhood interaction; I is the level of institutional influence. Y-axis: similarity between the state just before the event (1,000,000th iteration), and 100,000 iterations after the event. X-axis: event size as a percentage of the affected agents.

The overall results for partial and full content removal are comparable to those obtained for institutional structural damage in Figure 6. However, in this case, an inverse relation exists between number of cultural groups and similarity in the 32x32 scenarios for partial content removal; lower numbers of preexisting cultures here help in upkeeping similarity to previously existing cultural states. This relationship is not consistent for full content removal, and not established for the larger-sized grids.

Observing the gradient of the lines in the uniformly distributed events, *partial content removal* seems to have a larger impact at smaller sizes than *full content removal*. This is possible because partially destroying content of institutions could degrade the cultural border in the manner of a two-way street, making agents more likely to inherit traits from institutions of adjacent cultural groups. *Full content removal* might instead enable individuals to consistently rebuild their institutions based on their own cultural affiliation and make adoption of foreign institutional traits less likely, therefore preserving their pre-event state better.

**Institutional Conversion**

Figure 8 shows the effects of uniform and normally distributed *institutional conversion* on the scenarios as before, with partial conversion (top) and full content conversion (bottom).

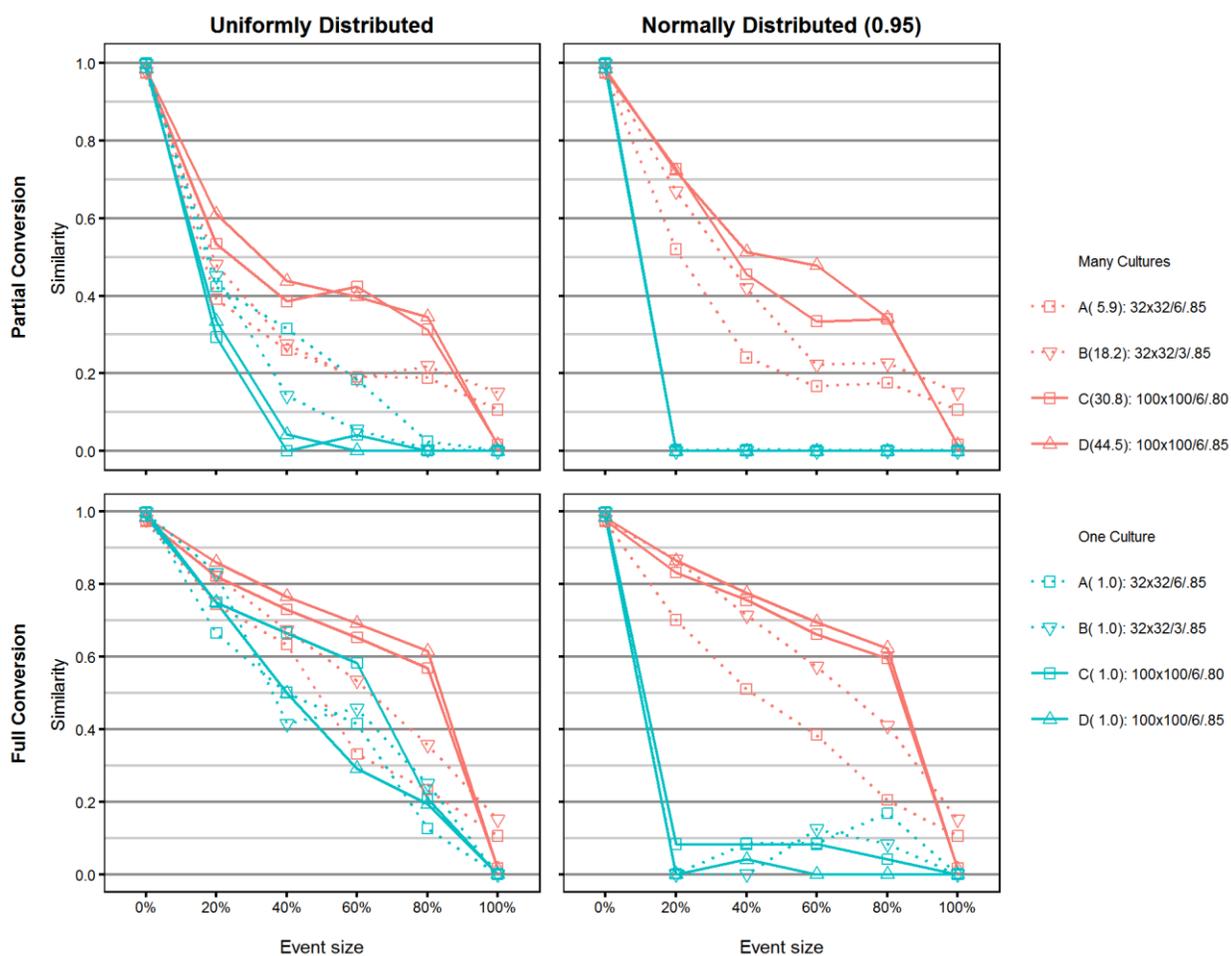

*Figure 8. Effects of partial and full institutional conversion.* Legend: S(G): NxN/R/I where S is the identifier; G is the average number of cultural groups; NxN is the number of rows and column; R is the distance for neighborhood interaction; I is the level of institutional influence. Y-axis: similarity between the state just before the event (1,000,000[th] iteration), and 100,000 iterations after the event. X-axis: event size as a percentage of the affected agents.

Both partial and full institutional conversion to foreign traits impact cultural composition of simulated societies post-event strongly, even at small levels, having the highest impact of all events: changing institutional content partially at 20% results in similarity of at best 60%-70%, while similar effects are reached for full conversion at around 30%.

As before, the distribution of the event, and cultural groups pre-event play a role in how societies develop. The here presented diverse societies show higher levels of resilience against institutional content conversation; moreover, the positive correction for 32x32 grid scenarios is now also present in the 100x100 scenarios (in all scenarios, the red triangles indicate higher similarity than the red squares). In contrast, monocultures tend to change quickly and completely at even low levels of institutional conversion. Only the scenario with full conversion at a uniform distribution presents a linear relation to the size of the event. Overall, across all cultures, a partial conversion is even more drastic than a full content conversion, similar to results from institutional content removal.

## Discussion

### Diversity and resilience

The primary research question of the here presented investigation was whether cultural diversity would be more resilient than monoculture. Across our results, we find consistent evidence of the simulated diverse societies being more resilient to events and catastrophes than monoculture societies, with a trend of more diversity increasing resilience for some specific scenarios.

A few exceptions are noticeable in the graphs. Most of these exceptions occur only for small event sizes; for example in the decimation, the apostasy, and the uniformly distributed foreigner events, the simulated monocultures remain completely intact (similarity=1), while the diverse simulated societies suffer some degree of change. This change is due to the constant social influence mechanism occurring between agents across neighboring cultural groups, which do not exist in monocultures. Even in the absence of an event (i.e. when event size is 0% on the X-axis) changes still occur in the diverse scenarios, and the magnitude of change remains a very similar level for, at least, 20% event sizes.

There is another set of isolated instances where monoculture was found to be on exception more resilient than diversity. Although the differences are small, it suggests that monoculture is advantageous in very particular circumstances given that the instances appear scattered (e.g. uniformly distributed partial conversion for scenarios A for event sizes between 20% and 40%).

Among some diverse scenarios, namely institutional conversion, institutional destruction, and to some extent, for apostasy, we observe a positive relation between the number of pre-existing cultural groups and their resilience: the larger the number of groups pre-event, the higher the similarity post-event. This relation is inverted for partial institutional content removal, and mixed results are observed in foreigner-related events. While the methodology supports analysis of this relation, and we chose scenarios that produced different number of groups (G) (ordered incrementally from A to D according to G (see Table 4)), in order to control for the number of groups, we had to adjust the parameters of the model: grid size, radius, and institutional influence. Any observation regarding the number of groups is confounded with at least one of those parameters. At the same time, it is clear that an indivisible dependency seems to exist between the number of groups and the parameters that create them, i.e. diversity is not artificially introduced, but a result of a complex process. It is not unlikely that this relationship (and other, unobserved, more complex ones) exist in real life, differentiating the emergence of diversity from monocultures. In other words, although it is not wrong to attribute the effect to the parameters, one can then argue that those parameters also produce more diversity, i.e. shifting the causality does not eliminate the correlation. Future work should attempt to vary the parameters in different ways to generate similar diversity levels while controlling for the parameters and events in a more stringent manner to better understand the relation between diversity, resilience, as well as neighborhoods and institutional influence, to uncover further relations that might not appear in the present paper.

**Event type, event intensity and resilience**

A second research question was the predicted link between large-scale and catastrophic events, their intensity (distribution and size) and how much they impact the cultural composition of societies in our simulation. The obtained results show a strong main effect of the intensity of the event, as well as qualitative differences between the events; meaning that some of the simulated cultural systems – and their differing parameters – are better equipped to deal with certain events than others.  For example, decimation by itself did not affect the composition of the here presented societies strongly except at very high levels: decimation rates at above 80% were necessary to achieve this. This is in line with some evidence found regarding mortality rates and their effects on societies in the past; for example, mortality rates as high as 90% have been cited for Maya populations due to violence and disease brought by the Spanish arrival (Coe 1999), and yet cultural diversity and strong cultural ties and institutions dated pre-invasion are abundant among modern Maya groups. As an example, there currently exist around 30 unique indigenous languages that are indisputably non-Spanish (Fischer and Brown 1997).

The introduction of foreigners was much more impactful: the simulated societies changed even at relatively small percentages of foreign outsiders or settlements. The data here suggests the survival of the foreign culture, which stabilizes, becoming part of society and its new composition. We also find indications for the proportionality of foreigners introduced compared to change in similarity observed. In the diverse scenarios and normally distributed events (Figure 5 – two scenarios of the right column), the impact is indeed proportional - at an introduction of 20% foreigners, the culture changes by roughly 20%. Small percentage differences between settlement and outsiders is due to the cohesive effects of the institutions in the former (Ulloa, Kacperski, and Sancho 2016). For uniformly distributed foreigners, major effects occur only upwards of sizes of 40% for settlers and 60% for outsiders. These numbers suggest that at least in our simulations, societies are very resilient at preserving their culture, even if foreigners share the same physical location, or if there is an institution that represents them. While great caution needs to be taken interpreting findings in the context outside the simulations, at least the here found data points to the fact that fears of a major loss of local traditions and institutions due to the presence of foreigners cannot be so easily supported; rather, especially at smaller event sizes, integration of foreigners is more probable. As a reference of modern day migration sizes, Germany holds a 23.6% non-citizen population, of which the two major migrant groups are Turkish (at 3.2%) and Polish (at 2.5%) (Destatis 2019).

Several events, namely full and partial institutional content removal, institutional damage and apostasy, never completely affected the cultural composition of diverse simulated societies; even at 100% event size the similarity remained above 0.5. This suggests that societies are able to rebuild a large part of their institutions from agents' features and interactions in order to prevent total disintegration. A stronger threat to the cultural composition of societies is institutional conversion, in which new traits are injected into old institutions. In all scenarios, cultural composition was heavily affected by institutional conversion, reinforcing the evidence of the important role that institutions have in up-keeping the stability of groups. Remarkably, this event was the strongest, even though no mechanism was employed to transmit the new traits from the institutions to the agents. Therefore, the effect stems only from the loss of institutional influence which debilitates the cultural borders between adjacent groups, i.e., neighboring agents that belonged to different groups were more likely to share traits as they no longer identified strongly with their institutions. The permeation of the catholic religion through native institutions in America during the colonization periods is one example of how institutions can be targeted by the introduction of foreign traits into societies, and therefore majorly change the cultural composition of a society (Koschorke et al. 2007; O'Connor 1989).

The data additionally shows an interesting pattern emerging from the institutional conversion event, which can also be observed in the content removal event: full institutional events are less effective in changing the overall cultural composition of a society than are partial institutional events. In the former, all traits of a selected institution are affected, whereas in the latter, only a percentage of traits are affected in all selected institutions. Although the (expected) total of affected traits is the same, the relation between the event size and the similarity is very different (in particular for diverse scenarios). The intuitive hypothesis would be that changing an institution wholly would affect agents more gravely; nevertheless, in effect, partial institutional events have more leverage over the population due to the way agents are influenced by institutions according to their similarity with each other (Equation (3)). Although a complete change of all traits in an institution removes the institutional influence over its agents, it is possible that a neighboring institution preserves all its traits, helping with the preservation of the culture. Conversely, a change in only some traits of neighboring institutions weakens the institutional influence onto all groups (included in the event distribution), and therefore erodes the bilateral relation between each pair of neighboring groups. Extending the previous example, Catholicism was relatively quickly wide-spread in the Americas due to the utilization of existent structures and institutions that were partially converted to Catholicism (Koschorke et al. 2007; Lafaye 1987); as a result there are an extensive number of syncretic creeds in America (González and González 2007). The results show how a partial conversion more effectively weakens the current cultural composition potentially making the population more susceptible to spreading new ideas.

**Applications and future directions**

Although we have given some real-life examples across this paper that fit observed patterns in the simulations for illustrative purposes, theoretical and explorative findings such as those presented here should be hesitantly applied in practice, and need definitely be validated with clear a priori hypotheses and confirmatory analyses. Indeed, the presented methodology allows for such analyses to be conducted, and combined events can be generated with CulSim and combined with additional parameters such as democracy and propaganda. Selected values should then be chosen carefully based on indicators from historical or archeological literature and tangible evidence wherever possible.

With this caveat, some generalized statements might provide impetus for finding further applicative purposes, considering the size and reliability of the effects. One, for societies as they are modelled here (relatively high institutional influence, relatively high connectivity), natural disasters (as long as they leave institutions largely unaffected) seem to play a lesser role in cultural change over time than do man-made disasters like wars and colonization, or events such as sudden changes in governance. Two, based on the data generated here, disaster relief following natural catastrophes or wars might harm a society's cultural composition if they include large-scale outside interference or partially affect the society's institutions; indeed, this finding seems in line with previous literature that indicates that the use of local resources and institutions to help stabilization in disaster regions is preferred (Neef and Shaw 2013), and that "international aid delivery system does not recognize the inherent conflicts of interest in existing community social structures" (Berke, Kartez, and Wenger 1993, 96). Increased care might allow for less disrupted rebuilding of local institutions from the pre-event population.

**Conclusion**

The here presented research proposes a means of analysis of cultural complexity and the resilience of cultural diversity in the face of large-scale events and catastrophes. The main issue has traditionally been the lack of a platform to experimentally test predictions, going beyond the so far empirical or philosophical approaches. Thus, its major contribution as a methodological paper is the employment of artificial societies to explore this sociocultural issue. CulSim (Ulloa et al, 2015) provides researchers with the flexible tool to explore concrete historical or future events by combining informationally conceptualized events, whose effects are analyzed in depth in the present paper. The approach taken in this paper is an adaptation of the model of cultural local convergence and global diversity by Axelrod (1997b) - using a recent extended model by Ulloa et al (2016), which includes institutions.

Thus far, not much consensus exists on which types of cultures are more resilient or which factors prevent societies from being changed or eradicated. The effect of a number of events such as decimation, foreign settlements, and institutional collapses on cultural composition of societies with different number of cultures were studied. The similarity of the culture pre-event and post-event was analyzed as the main outcome variable, while size of impact and variability of events provided additional information. Theoretically, we contribute to the literature on diversity as a mechanism of resilience. With the findings, we provide new insights into how societies are affected by large-scale events and catastrophes; this can allow predictions to be made about how to deal with devastating events while preserving the cultural composition of societies, leveraging diversity as one possible protective factor. Further research can extend the here presented model to study events such as modern wars, in which extensive numerical data exist, as a combination of the presented events.